\documentclass[%
 reprint,
nofootinbib,
 amsmath,amssymb,
 aps, showkeys,
]{revtex4-2}

\usepackage[english]{babel}

\usepackage{graphicx}
\usepackage{dcolumn}
\usepackage{bm}
\usepackage{txfonts}
\usepackage{mathtools}
\usepackage[colorlinks=true, allcolors=blue]{hyperref}
\usepackage{esint}
\usepackage{siunitx}
\usepackage{float}

\usepackage{wasysym}

\begin{document}
\selectlanguage{english}

\preprint{APS/123-QED}

\title{A No-go Theorem for Superposed Actions\\{\small (Making Schrödinger's Cat Quantum Nonlocal)}}

\author{Szymon Łukaszyk}
 \email{szymon@patent.pl}
\affiliation{%
 Łukaszyk Patent Attorneys, ul. Głowackiego 8, 40-052 Katowice, Poland
}%


\begin{abstract}
The Extended Wigner's Friend thought experiment, which involves a quantum system with an agent who draws conclusions based on the results of a measurement of a quantum state provided in two nonorthogonal versions by another agent, led its designers to the conclusion that quantum theory cannot consistently explain the use of itself. It has also been suggested that this thought experiment is equivalent to entangled state (Bell-type) experiments. This study indicates that the assumption of the first Wigner's friend's freedom of choice, regarding how to prepare a quantum state in one of the two available nonorthogonal versions, invalidates such equivalence. A no-go theorem for superposed actions is derived on this basis. It is also argued that modeling Wigner-type experiments under the principle of locality, i.e., using enclosed containers modeled as composite, many-body quantum states, is fundamentally wrong as it neglects quantum nonlocality.
\end{abstract}



\maketitle


\section{Introduction}\label{sec:1}
The Extended Wigner's Friend (EWF) thought experiment appears to contain a contradiction, which initially led its designers to abandon the idea of a single world \cite{1}, and afterward to reach the conclusion that Quantum Theory (QT) cannot consistently describe the use of itself \cite{2}, because the contradiction arrives within a single world. Indeed, the many-worlds interpretation of QT has its own paradoxical features \cite{3}, and it is not only counterfactually indefinable but also factually indefinite \cite{4}. On the other hand, if QT cannot consistently describe the use of itself, it would be an ultimate theory for the perceived nature, since no consistent system of axioms can demonstrate its own consistency, which is the conclusion of the 2$^\text{nd}$ Gödel's incompleteness theorem if only the Peano axioms of arithmetic are universally valid (i.e., as long as the perceived nature is subject to these axioms).

It is, however, questionable how QT is used in the EWF to describe the use of itself. The EWF is allegedly not just a thought experiment. Its authors claim, for example, that if the EWF were implemented as a game between a gambler and a casino, both parties would likely have entered into a dispute, putting forward contradicting assertions based on quantum-mechanical reasoning. These assertions would have to be accepted as two alternative (observer-dependent) facts (defined as observations) about what was the result of the first measurement in this thought experiment \cite{2}.

To date, no consensus has been achieved in the scientific community regarding the universal validity of any particular interpretation of the measurement problem \cite{5}. The problem manifests in various Wigner-type experiments \cite{6,7}, including the EWF \cite{2}, and has been highly debated \cite{8,9,10,11,12,13,14,15,16,17}.

The paper is structured as follows. Section \ref{sec:2} outlines the EWF setup. Section \ref{sec:3} introduces the concept of a superposed action and presents a no-go theorem for its implementation. Section \ref{sec:4} discusses the findings of the paper and the feasibility of implementing Bell-Wigner-type experiments (e.g., \cite{19,20,21,22}) involving entangled states, whereas Section \ref{sec:5} concludes the results of this study.

\section{THE EWF WITH A SUPER-OBSERVER}\label{sec:2}
In each round of the EWF, in her sealed lab Alice prepares a first state (a qubit)
\begin{equation}\label{1}
\left|\alpha\right\rangle=\sqrt{\frac{1}{3}}\left|\text{h}\right\rangle + \sqrt{\frac{2}{3}}\left|\text{t}\right\rangle,
\end{equation}
measures it at a certain time $t_0$ in a basis \{$\left|\text{h}\right\rangle$, $\left|\text{t}\right\rangle$\}, records the measurement, and prepares a second state
\begin{equation}\label{2}
\left|\beta\right\rangle=\left\{
\begin{array}{ll}
\left|0\right\rangle &\, \text{if} \, \left|\alpha\right\rangle=\left|\text{h}\right\rangle\\ \sqrt{\frac{1}{2}}\left(\left|0\right\rangle + 
\left|1\right\rangle\right) &\, \text{if} \, \left|\alpha\right\rangle=\left|\text{t}\right\rangle 
\end{array}
\right..
\end{equation}

Alice hands the second state\footnote{Describing (gruesomely) this preparation process: Alice puts a cat into a Schrödinger's box provided with all the necessary equipment (a Geiger counter, a flask of poison, etc.), if she measures $\left|t\right\rangle$, but she poisons the cat before putting its corpse into the box, if she measures $\left|h\right\rangle$. She delivers the box to Bob, who will open it to find out whether the cat is alive ($\left|1\right\rangle$) or poisoned ($\left|0\right\rangle$).} \eqref{2} to Bob residing in another sealed lab. At a later time $t_1 > t_0$, Bob measures the received second state in a basis \{$\left|0\right\rangle$, $\left|1\right\rangle$\}. At an even later time $t_2 > t_1$ Charlie measures the first state \eqref{1} emitted from Alice's lab in a first Hadamard basis defined as
\begin{equation}\label{3}
\left|\square\right\rangle \doteq
\sqrt{\frac{1}{2}}\left(\left|\text{h}\right\rangle + \left|\text{t}\right\rangle\right),\quad \left|\bigcirc\right\rangle \doteq \sqrt{\frac{1}{2}}\left(\left|\text{h}\right\rangle - \left|\text{t}\right\rangle \right),
\end{equation}
and the second state \eqref{2} emitted from Bob's lab in a second Hadamard basis defined as
\begin{equation}\label{4}
\left|+\right\rangle \doteq \sqrt{\frac{1}{2}}\left(\left|0\right\rangle + \left|1\right\rangle\right),\quad \left|-\right\rangle \doteq \sqrt{\frac{1}{2}}\left(\left|0\right\rangle - \left|1\right\rangle \right),
\end{equation}
and the round is completed. Since Charlie does not know the measurement results of the first \eqref{1} and the second \eqref{2} states emitted from Alice's and Bob's labs, statistics of Charlie's measurements of the first state \eqref{1} in a large number of rounds of this experiment can be described by the mixed state density matrix
\begin{equation}\label{5}
\rho_{\alpha}=\frac{1}{3}\left|\text{h}\right\rangle \left\langle \text{h} \right| + \frac{2}{3}\left|\text{t}\right\rangle \left\langle \text{t} \right|,
\end{equation}
while statistics of Charlie's measurements of the second state \eqref{2}, by the mixed state density matrix
\begin{equation}\label{6}
\rho_{\beta}=\frac{1}{3}\left|0\right\rangle \left\langle 0 \right| + \frac{2}{3}\left|+\right\rangle \left\langle + \right|,
\end{equation}
which has larger purity $(\operatorname{trace}\left(\rho_{\beta}^2 \right) = 7/9)$ than purity $(\operatorname{trace}\left(\rho_{\beta_1}^2 \right) = 5/9)$ of the density matrix
\begin{equation}\label{7}
\rho_{\beta_1}=\frac{2}{3}\left|0\right\rangle \left\langle 0 \right| + \frac{1}{3}\left|1\right\rangle \left\langle 1 \right|,
\end{equation}
that would yield the same measurement probabilities.

So far, there is nothing contradicting (as nothing is entangled) in this setup. Measurement probabilities for a large number of rounds are nonzero 
$p\left(\left|\text{h}\right\rangle\right) = p\left(\left|1\right\rangle\right) = 1/3$, 
$p\left(\left|\text{t}\right\rangle\right) = p\left(\left|0\right\rangle\right) = 2/3$, 
$p\left(\left|\square\right\rangle\right) = p\left(\left|\bigcirc\right\rangle\right) = 1/2$, 
$p\left(\left|+\right\rangle\right) = 5/6$, 
$p\left(\left|-\right\rangle\right) = 1/6$.

Due to the principle of locality, however, these kinds of thought experiments (Wigner-type experiments) involving a qubit and someone who measures this qubit contained in a box or a lab isolated from the environment (Schrödinger's cat, Wigner's friend \cite{6}, Deutsch's variant \cite{7}, and the EWF) are described as coherent \emph{big quantum states} of those boxes or labs. In the case of Alice's lab, for example, this big quantum state is described as a tensor product of the basic quantum state \eqref{1}, of some device that enables measurement of this basic state and finally of Alice herself in her lab including her sense organs, her brain, etc. This indeed seems unrealistic, even if not explicitly precluded by the laws of QT as such. Schrödinger's cat has nothing to do with quantum information science, even if the latter can be harnessed to kill the cat. Deutsch's variant \cite{7}, in which a friend informs Wigner that she has a definite measurement result, but does not reveal this result, so as not to accidentally destroy the superposition of the \emph{big quantum state} from inside the lab, is particularly instructive.

Even if the assumptions (Q, C, S) of the EWF do not explicitly include locality \cite{8}, the locality is used in the EWF to model the enclosed immediate surroundings of Alice's and Bob's labs including Alice and Bob themselves and their actions in space and time, as a coherent, entangled \emph{big quantum state}. Following such a \emph{box-thinking}, some authors assume that Charlie is allowed to measure either the entire Alice's and Bob's labs, thus, \emph{erasing} [\emph{sic}] Alice's and Bob's records\footnote{The author can hardly imagine someone measuring (\emph{observing}) the enclosed room in which he's writing this footnote and erasing the author's record of this fact due to this \emph{observation}.}, or \emph{preserve the records} by measuring only a part of a lab \cite{9}.

The \emph{big quantum state} of the EWF can be defined \cite{10} by Charlie as
\begin{equation}\label{8}
\left|\psi_{AB}\right\rangle =
\sqrt{\frac{1}{3}}\left|\text{h}0\right\rangle +
\sqrt{\frac{1}{3}}\left|\text{t}0\right\rangle +
\sqrt{\frac{1}{3}}\left|\text{t}1\right\rangle,
\end{equation}
after excluding Alice, Bob, and their devices, the exact properties of which do not need to be specified (\cite{19}, p. 16). They correspond to the definite-measurement-result information in Deutsch's variant \cite{7} of the Wigner-type experiment and are the same, regardless of the measurement result. Nonmaximally entangled state \eqref{8} shows that, unlike other Wigner-type experiments, only the EWF pretends to be a Bell-type one.

In Charlie's bases \{$\left|\square\right\rangle$, $\left|\bigcirc\right\rangle$\} and \{$\left|+\right\rangle$, $\left|-\right\rangle$\} the state \eqref{8} is
\begin{equation}\label{9}
\begin{split}
\left|\psi_{C}\right\rangle =&
\sqrt{\frac{9}{12}}\left|\square+\right\rangle +
\sqrt{\frac{1}{12}}\left|\square-\right\rangle -\\&
\sqrt{\frac{1}{12}}\left|\bigcirc+\right\rangle +
\sqrt{\frac{1}{12}}\left|\bigcirc-\right\rangle =\\&
\sqrt{\frac{9}{12}}\left|\square+\right\rangle +
\sqrt{\frac{1}{12}}\left|\square-\right\rangle-
\sqrt{\frac{2}{12}}\left|\bigcirc1\right\rangle,
\end{split}
\end{equation}
and in mixed bases 
\{$\left|\text{h}\right\rangle$, $\left|\text{t}\right\rangle$\},
\{$\left|0\right\rangle$, $\left|1\right\rangle$\},
\{$\left|\square\right\rangle$, $\left|\bigcirc\right\rangle$\},
\{$\left|+\right\rangle$, $\left|-\right\rangle$\} it is
\begin{equation}\label{10}
\begin{split}
\left|\psi_{AC}\right\rangle = &\sqrt{\frac{1}{6}}\left|\text{h}+\right\rangle + \sqrt{\frac{1}{6}}\left|\text{h}-\right\rangle + \sqrt{\frac{4}{6}}\left|\text{t}+\right\rangle =\\&
\sqrt{\frac{1}{3}}\left|\text{h}0\right\rangle + \sqrt{\frac{2}{3}}\left|\text{t}+\right\rangle,
\end{split}
\end{equation}
\begin{equation}\label{11}
\begin{split}
\left|\psi_{CB}\right\rangle =&
\sqrt{\frac{4}{6}}\left|\square0\right\rangle + 
\sqrt{\frac{1}{6}}\left|\square1\right\rangle - 
\sqrt{\frac{1}{6}}\left|\bigcirc1\right\rangle =\\&
\sqrt{\frac{2}{3}}\left|\square0\right\rangle + 
\sqrt{\frac{1}{3}}\left|\text{t}1\right\rangle.
\end{split}
\end{equation}

The following simple argumentation used to expose the contradiction of the EWF \cite{10} is similar to the one used in Ref. \cite{23} to illustrate the mystery of the quantum cakes (“a simple 'real-world' explanation of the proof of quantum mechanical nonlocality without the use of inequalities”; We will further call it “quantum-cakes explanation”):
\begin{enumerate}
  \item[1. ] We know from $\left|\psi_C\right\rangle$ \eqref{9} that measurements of $\left|\bigcirc\right\rangle$ and $\left|-\right\rangle$ are possible with probability $p = 1/12$;
  \item[2. ] We know from $\left|\psi_{AC}\right\rangle$ \eqref{10} that $\left|-\right\rangle \Rightarrow \left|\text{h}\right\rangle$\footnote{Of course the notation  “$\left|-\right\rangle \Rightarrow \left|\text{h}\right\rangle$" means that measurement of $\left|-\right\rangle$ implies measurement of $\left|\text{h}\right\rangle$, etc.};
  \item[2'.] We know from $\left|\psi_{CB}\right\rangle$ \eqref{11} that $\left|\bigcirc\right\rangle \Rightarrow \left|1\right\rangle$; and
  \item[3. ] We know from $\left|\psi_{AB}\right\rangle$ \eqref{8} that $\left|\text{h}\right\rangle \Rightarrow \left|0\right\rangle$.
\end{enumerate}
In other words, the state $\left|\psi_{C}\right\rangle$ \eqref{9} measured as $\left|\bigcirc-\right\rangle$ imply the term $\left|\text{h}1\right\rangle$, absent in the expansion $\left|\psi_{AB}\right\rangle$ \eqref{8}, which is contradicting.

We note in passing that the state \eqref{8} in bases \eqref{3}, \eqref{4} is more symmetric than the oven state and bases used in Ref. \cite{23}, and that the concept of time is irrelevant in the quantum-cakes explanation.

The EWF contradiction is, however, derived in Ref. \cite{2} on the grounds of the following predictions made by Alice, Bob, and Charlie at different times $t_0$, $t_1$, and $t_2$ during a happy round of this experiment. Each round goes like this:
\begin{enumerate}
  \item[(0)] Alice and Bob prepare and/or measure their states \eqref{1} and \eqref{2} according to the procedure described at the outset, and Charlie measures $\left|\psi_C\right\rangle$, which describes the \emph{big quantum state} of both Alice's and Bob's labs (at time $t_2$);
  \item[$\frownie$] if Charlie's measurement result is $\left|\square+\right\rangle$, $\left|\square-\right\rangle$, or $\left|\bigcirc+\right\rangle$, the round is completed with no contradiction, and a new round begins;
  \item[$\smiley$] if Charlie's measurement result is $\left|\bigcirc-\right\rangle$, then they have a happy round and
  \begin{enumerate}
    \item[(A)] Alice knows from the state \eqref{2} that $\left|\text{t}\right\rangle$ at time $t_0$ implies $\left|+\right\rangle$ at time $t_2$;
    \item[(B)] Bob knows from the state \eqref{2} that $\left|1\right\rangle$ at time $t_1$ implies $\left|\text{t}\right\rangle$ at time $t_0$; and
    \item[(C)] Charlie knows from the state $\left|\psi_C\right\rangle$ \eqref{9} that $\left|\bigcirc\right\rangle$ at time $t_2$ implies $\left|1\right\rangle$ at time $t_1$.
  \end{enumerate}  
\end{enumerate}
All these four conditions, in a happy round of the EWF, also contradict each other: if $\smiley$ then $\left|\bigcirc\right\rangle \Rightarrow \left|1\right\rangle$ (C), $\left|1\right\rangle \Rightarrow \left|\text{t}\right\rangle$ (B), and $\left|\text{t}\right\rangle \Rightarrow \left|+\right\rangle$ (A). 
In other words, if $\smiley$ then $(\text{C} \Rightarrow \text{B}) \cap (\text{B} \Rightarrow \text{A}) \cap (\text{A} \Rightarrow \neg\text{C})$, so $\text{C} \Rightarrow \neg\text{C}$ which is contradicting.

We will further call this argumentation “superposed-action explanation”. It is grounded on the assumption that Alice's and Bob's states \eqref{1} and \eqref{2} have evolved unitarily to a composite entangled state \eqref{8} after Alice handed the second state to Bob.

\section{SUPERPOSED ACTION}\label{sec:3}
Let us have a closer look at how the state \eqref{8} could possibly be created using the procedure of the EWF under the standard assumptions of Wigner's friend thought experiments. Namely, it is assumed that from Charlie's super-observer perspective, after the state \eqref{1} is measured by Alice at time $t_0$, it becomes a tensor product
\begin{equation}\label{12}
\begin{split}
\left|\alpha(t_0)\right\rangle =& \sqrt{\frac{1}{3}}\left|\text{h}\right\rangle \otimes \left|\text{Alice knows h}\right\rangle+\\&
\sqrt{\frac{2}{3}}\left|\text{t}\right\rangle \otimes \left|\text{Alice knows t}\right\rangle.
\end{split}
\end{equation}

For the next stage of the EWF, a usual definition of the
\begin{center}
\textbf{Freedom of Choice}. The choice of measurement settings is statistically independent from the rest of the experiment.
\end{center}
(Statement 3 in the no-go theorem of Ref. \cite{20}), seems, however, insufficient. Probability amplitudes are fixed in all the states and bases of the EWF (they do not need to be chosen in each round of this thought experiment). Therefore, the proposed, amended definition is:
\begin{center}
\textbf{Freedom of Choice}. The observer's opportunity and autonomy to prepare a quantum state in one of the two available nonorthogonal versions, wherein that state will be measured later by (an)other observer(s), is statistically independent from the rest of the experiment.
\end{center}
There is no guarantee that an agent inside the lab will act as expected. And that makes it impossible to derive arguments that implicitly assume that she acts properly and reports valid measurement results.

We note in passing that such an amended Freedom of Choice assumption is not explicitly made in \cite{2}. It is therefore another hidden assumption of this thought experiment among other hidden assumptions \cite{8}.

Therefore, the following tensor product, corresponding to the tensor product \eqref{12}
\begin{equation}\label{13}
\begin{split}
&\left|\beta(t_1)\right\rangle \ne\\
&\sqrt{\frac{1}{3}}\left|0\right\rangle	\otimes\\
&\left|\text{Bob knows $0$, as Alice knew h and prepared $\left|0\right\rangle$ at $t_0$}\right\rangle+\\
&\sqrt{\frac{1}{3}}\left|0\right\rangle	\otimes\\
&\left|\text{Bob knows $0$, as Alice knew t and prepared $\left|+\right\rangle$ at $t_0$}\right\rangle+\\
&\sqrt{\frac{1}{3}}\left|1\right\rangle	\otimes\\
&\left|\text{Bob knows $1$, as Alice knew t and prepared $\left|+\right\rangle$ at $t_0$}\right\rangle
\end{split}
\end{equation}
cannot be valid for Bob measuring, at time $t_1$, the state \eqref{2} prepared by Alice at time $t_0$, as this would violate Alice's freedom of choice. The absence of observer-independent observations (facts) \cite{19,20} allows one to discuss superpositions of observer-dependent measurements \eqref{12} of time-independent quantum states, but not to discuss superpositions of observer-dependent and time-dependent observer's actions \eqref{13}.

Unlike Alice's measurements (observations), Alice's actions are relative not only to Alice \cite{19,20}; they can be measured (observed) also by Bob and Charlie.

We are now ready to state the

\textbf{Theorem} (A No-Go Theorem For Superposed Actions):
There is no unitary transformation that would transform a finite-dimensional, pure quantum state and at least two nonorthogonal versions of a finite-dimensional, mixed quantum state into an entangled, pure quantum state. Such a transformation would not preserve the inner product and thus would not be unitary.
In other words, there is no unitary transformation that would transform a finite-dimensional, pure quantum state and at least two nonorthogonal versions of a finite-dimensional, mixed quantum state prepared within an enclosed volume due to some agent actions, into an entangled, pure quantum state that would describe this volume as a composite, many-body entangled quantum system \cite{10}. Such a transformation would additionally neglect quantum nonlocality.
This theorem applies to our case, where the two nonorthogonal versions of the finite-dimensional, mixed quantum state are prepared by Alice's actions and then this impure superposition is measured by Bob.

\textbf{Proof}: Without loss of generality, assume that the pure quantum state is the state \eqref{1}, two nonorthogonal versions of a mixed quantum state are represented by \eqref{2}, while the entangled, pure quantum state is represented by \eqref{8}.
From Charlie's super-observer perspective quantum register containing the first qubit \eqref{1} and the second state $\left|\beta\right\rangle = \left|0\right\rangle$ or $\left|\beta\right\rangle = (\left|0\right\rangle + \left|1\right\rangle)/\sqrt{2}$ after it was prepared by Alice would initially contain either two separable pure states 
$\left|\text{h}0\right\rangle$ and $\left|\text{t}0\right\rangle$ (if Alice measured $\left|\text{h}\right\rangle$ at time $t_0$)
\begin{equation}\label{14}
\begin{split}
\left|\psi_{\text{h}0}\right\rangle =& \left(\sqrt{\frac{1}{3}}\left|\text{h}\right\rangle + \sqrt{\frac{2}{3}}\left|\text{t}\right\rangle\right)\otimes \left|0\right\rangle=\\
&\sqrt{\frac{1}{3}}\left|\text{h}0\right\rangle + \sqrt{\frac{2}{3}}\left|\text{t}0\right\rangle
\end{split}
\end{equation}
or four separable pure states $\left|\text{h}0\right\rangle$, $\left|\text{h}1\right\rangle$, $\left|\text{t}0\right\rangle$, and $\left|\text{t}1\right\rangle$ (if Alice measured $\left|\text{t}\right\rangle$ at time $t_0$):
\begin{equation}\label{15}
\begin{split}
\left|\psi_{\text{t}01}\right\rangle =& 
\left(\sqrt{\frac{1}{3}}\left|\text{h}\right\rangle + 
\sqrt{\frac{2}{3}}\left|\text{t}\right\rangle\right) 
\otimes \left(\sqrt{\frac{1}{2}}\left|0\right\rangle + 
\sqrt{\frac{1}{2}}\left|1\right\rangle\right)=\\&
\sqrt{\frac{1}{6}}\left|\text{h}0\right\rangle + 
\sqrt{\frac{1}{6}}\left|\text{h}1\right\rangle +
\sqrt{\frac{2}{6}}\left|\text{t}0\right\rangle +
\sqrt{\frac{2}{6}}\left|\text{t}1\right\rangle.
\end{split}
\end{equation}

The reader might be tempted at this point to question the validity of equations \eqref{14} and \eqref{15}. If the first qubit is measured (by Alice) it retains its state and instead of \eqref{14} and \eqref{15} we would have respectively $\left|\text{h}0\right\rangle$ or $\left|\text{t}\right\rangle\otimes(\left|0\right\rangle + \left|1\right\rangle)/\sqrt{2} = \left|\text{t}+\right\rangle$. The point here is that we consider three versions of the EWF. The 1$^\text{st}$ one deals with just two states \eqref{1} and \eqref{2} that are not entangled and retain their states after measurements. This version is described by density matrices \eqref{5}, \eqref{6} showing no contradiction. The 2$^\text{nd}$ one \cite{1,2,10} assumes the entanglement of these states along with the \emph{content} of Alice's lab. Although the author considers such an assumption unrealistic, it is \emph{a priori} assumed to be correct. However, in this setup, Alice's action must be taken into account, as Bob's lab cannot be described by the state \eqref{13}, which leads us to Equations \eqref{14}, \eqref{15}, and the 3$^\text{rd}$ version of the EWF.

It is straightforward to see that the inner products of the states \eqref{14} and \eqref{15} with \eqref{8} are different
\begin{equation}\label{16}
\left\langle\psi_{\text{h}0}|\psi_{AB}\right\rangle=\frac{1+\sqrt{2}}{3} \approx 0.8047,
\end{equation}
\begin{equation}\label{17}
\left\langle\psi_{\text{t}01}|\psi_{AB}\right\rangle=\frac{1+2\sqrt{2}}{3\sqrt{2}} \approx 0.9024,
\end{equation}
whereas
\begin{equation}\label{18}
\left\langle\psi_{\text{h}0}|\psi_{\text{t}01}\right\rangle=\frac{1}{\sqrt{2}} \approx 0.7071,
\end{equation}
which completes the proof.

In order to affect the unitary evolution of such two different quantum registers to bring them both to the entangled state $\left|\psi_{AB}\right\rangle$ \eqref{8}, that would be the same regardless of the initial state $\left|\psi_{\text{h}0}\right\rangle$ \eqref{14} or $\left|\psi_{\text{t}01}\right\rangle$ \eqref{15}, Alice must use two different variants of some suitable $4\times4$ unitary matrices.

If she measures the state \eqref{1} as $\left|\text{h}\right\rangle$ she may first use the following unitary matrix $A_{\text{h}0}$
\begin{equation}\label{19}
A_{\text{h}0}\left|\psi_{\text{h}0}\right\rangle = \begin{bmatrix}
\sqrt{\frac{1}{3}} & 0                  & \sqrt{\frac{2}{3}}    & 0\\
0                  & \sqrt{\frac{1}{3}} & 0                     & -\sqrt{\frac{2}{3}} \\
\sqrt{\frac{2}{3}} & 0                  & -\sqrt{\frac{1}{3}}   & 0\\
0                  & \sqrt{\frac{2}{3}} & 0                     & \sqrt{\frac{1}{3}}
\end{bmatrix}
\begin{bmatrix} 
\sqrt{\frac{1}{3}}\\0\\\sqrt{\frac{2}{3}}\\0
\end{bmatrix} =
\begin{bmatrix}
1\\0\\0\\0
\end{bmatrix} = \left|\psi_{00}\right\rangle,
\end{equation}
while if she measures the state \eqref{1} as $\left|\text{t}\right\rangle$ she may first use the following unitary matrix $A_{\text{t}01}$
\begin{equation}\label{20}
A_{\text{t}01}\left|\psi_{\text{t}01}\right\rangle = \begin{bmatrix}
 \sqrt{\frac{1}{6}} &  \sqrt{\frac{1}{6}} &  \sqrt{\frac{1}{3}} &  \sqrt{\frac{1}{3}} \\
 \sqrt{\frac{1}{3}} &  \sqrt{\frac{1}{3}} & -\sqrt{\frac{1}{6}} & -\sqrt{\frac{1}{6}} \\
-\sqrt{\frac{1}{6}} &  \sqrt{\frac{1}{6}} &  \sqrt{\frac{1}{3}} & -\sqrt{\frac{1}{3}} \\
 \sqrt{\frac{1}{3}} & -\sqrt{\frac{1}{3}} &  \sqrt{\frac{1}{6}} & -\sqrt{\frac{1}{6}}
\end{bmatrix}
\begin{bmatrix} 
\sqrt{\frac{1}{6}} \\ \sqrt{\frac{1}{6}} \\ \sqrt{\frac{1}{3}} \\ \sqrt{\frac{1}{3}}
\end{bmatrix} =
\begin{bmatrix}
1\\0\\0\\0
\end{bmatrix} = \left|\psi_{00}\right\rangle.
\end{equation}
Then she may use the following unitary matrix $R$
\begin{equation}\label{21}
R\left|\psi_{00}\right\rangle = \begin{bmatrix}
\sqrt{\frac{1}{3}} & 0 &  \sqrt{\frac{1}{2}} &  \sqrt{\frac{1}{6}} \\
0                  & 1 &  0                  &  0 \\
\sqrt{\frac{1}{3}} & 0 & -\sqrt{\frac{1}{2}} &  \sqrt{\frac{1}{6}} \\
\sqrt{\frac{1}{3}} & 0 &  0                  & -\sqrt{\frac{2}{3}}
\end{bmatrix}
\begin{bmatrix} 
1\\0\\0\\0
\end{bmatrix} =
\begin{bmatrix} 
\sqrt{\frac{1}{3}} \\ 0 \\ \sqrt{\frac{1}{3}} \\ \sqrt{\frac{1}{3}}
\end{bmatrix} = \left|\psi_{AB}\right\rangle,
\end{equation}
to receive $\left|\psi_{AB}\right\rangle = R \left|\psi_{00}\right\rangle$ from $\left|\psi_{00}\right\rangle = A_{\text{h}0} \left|\psi_{\text{h}0}\right\rangle$ or $\left|\psi_{00}\right\rangle = A_{\text{t}01} \left|\psi_{\text{t}01}\right\rangle$.

There are obviously infinitely many possibilities of unitary transformations that would bring $\left|\psi_{\text{h}0}\right\rangle$ or $\left|\psi_{\text{t}01}\right\rangle$ to $\left|\psi_{AB}\right\rangle$ \eqref{8}, as groups of unitary $4 \times 4$ matrices act transitively on the unit vectors in Hilbert spaces over the complex numbers ($\mathbb{C}^4$). The above exemplary forms of these matrices model the evolution of the EWF, provided they are correctly used.

Indeed, such an approach assumes that in order to arrive at $\left|\psi_{AB}\right\rangle$ Alice must after recording the outcome of her measurement of the first qubit to be either $\left|\text{h}\right\rangle$ or $\left|\text{t}\right\rangle$ act according to this outcome in a manner predefined by \eqref{2} by applying an appropriate unitary matrix transformation.

But what if Alice makes a mistake and applies a \emph{wrong} transformation $RA_{\text{h}0}$ or $RA_{\text{t}01}$, respectively to the quantum register $\left|\psi_{\text{t}01}\right\rangle$ \eqref{15} or $\left|\psi_{\text{h}0}\right\rangle$ \eqref{14}?
Then she will obtain the following two states

\begin{equation}\label{22}
\left|\psi_{AB\text{th}}\right\rangle=
0.847\left|\text{h}0\right\rangle +
0.514\left|\text{t}0\right\rangle - 
0.136\left|\text{t}1\right\rangle,
\end{equation} 
\begin{equation}\label{23}
\left|\psi_{AB\text{ht}}\right\rangle=
0.680\left|\text{h}0\right\rangle - 
0.236\left|\text{h}1\right\rangle +
0.680\left|\text{t}0\right\rangle - 
0.136\left|\text{t}1\right\rangle,
\end{equation} 
certainly with nonzero probability, if one assumes that Alice has the freedom of choice in affecting the unitary evolution of the initial state $\left|\psi_{\text{h}0}\right\rangle$ \eqref{14} or $\left|\psi_{\text{t}01}\right\rangle$ \eqref{15}. Alice's freedom of choice implies her fallibility. These states in mixed bases 
\{$\left|\text{h}\right\rangle$, $\left|\text{t}\right\rangle$\},
\{$\left|0\right\rangle$, $\left|1\right\rangle$\},
\{$\left|\square\right\rangle$, $\left|\bigcirc\right\rangle$\}, and
\{$\left|+\right\rangle$, $\left|-\right\rangle$\} \eqref{3}, \eqref{4}
are for $\left|\psi_{AB\text{th}}\right\rangle$
\begin{equation}\label{24}
\begin{split}
\left|\psi_{C\text{th}}\right\rangle=&
0.612\left|\square+\right\rangle +
0.749\left|\square-\right\rangle +\\&
0.235\left|\bigcirc+\right\rangle +
0.099\left|\bigcirc-\right\rangle,
\end{split}
\end{equation} 
\begin{equation}\label{25}
\begin{split}
\left|\psi_{AC\text{th}}\right\rangle=&
0.599\left|\text{h}+\right\rangle +
0.599\left|\text{h}-\right\rangle +\\&
0.267\left|\text{t}+\right\rangle +
0.460\left|\text{t}-\right\rangle,
\end{split}
\end{equation} 
\begin{equation}\label{26}
\begin{split}
\left|\psi_{CB\text{th}}\right\rangle=&
0.962\left|\square0\right\rangle +
0.236\left|\bigcirc0\right\rangle -\\&
0.096\left|\square1\right\rangle +
0.096\left|\bigcirc1\right\rangle,
\end{split}
\end{equation} 
and for $\left|\psi_{AB\text{ht}}\right\rangle$
\begin{equation}\label{27}
\begin{split}
\left|\psi_{C\text{ht}}\right\rangle=&
0.495\left|\square+\right\rangle +
0.866\left|\square-\right\rangle -\\&
0.050\left|\bigcirc+\right\rangle +
0.050\left|\bigcirc-\right\rangle,
\end{split}
\end{equation} 
\begin{equation}\label{28}
\begin{split}
\left|\psi_{AC\text{ht}}\right\rangle=&
0.315\left|\text{h}+\right\rangle +
0.648\left|\text{h}-\right\rangle +\\&
0.385\left|\text{t}+\right\rangle +
0.577\left|\text{t}-\right\rangle,
\end{split}
\end{equation} 
\begin{equation}\label{29}
\begin{split}
\left|\psi_{CB\text{ht}}\right\rangle=&
0.962\left|\square0\right\rangle -
0.263\left|\square1\right\rangle -\\&
0.070\left|\bigcirc1\right\rangle.
\end{split}
\end{equation} 

Quantum-cakes-explanation fails both for $\left|\psi_{AB\text{th}}\right\rangle$ and $\left|\psi_{AB\text{ht}}\right\rangle$ ($\ast$ denotes $\text{th}$ or $\text{ht}$):

\begin{enumerate}
  \item We know from $\left|\psi_{C\ast}\right\rangle$ \eqref{24}, \eqref{27} that measurements of $\left|\bigcirc-\right\rangle$ are possible; but
  \item We know from $\left|\psi_{AC\ast}\right\rangle$ \eqref{25}, \eqref{28} that 
the $\left|-\right\rangle$ outcome does not imply $\left|\text{h}\right\rangle$ outcome, unlike in $\left|\psi_{AC}\right\rangle$ \eqref{10}.
\end{enumerate}

Superposed-action explanation also fails, if Alice makes a mistake. Charlie can measure $\left|\psi_{C\ast}\right\rangle$ at time $t_2$ and observe $\left|\bigcirc-\right\rangle$ but Alice's claim based on \eqref{2} that $\left|\text{t}\right\rangle$ measured at time $t_0$ implies $(\left|0\right\rangle + \left|1\right\rangle)/\sqrt{2} = \left|+\right\rangle)$ measured at time $t_2$ is false merely due to her own mistake. Neither Bob's claim, also based on \eqref{2}, that $\left|1\right\rangle$ measured at time $t_1$ implies $\left|\text{t}\right\rangle$ measured at time $t_0$ is true.

Elementary arithmetic shows that any normalized nonmaximally entangled state
\begin{equation}\label{30}
\left|\psi_{AB}\right\rangle =
a\left|\text{h}0\right\rangle +
c\left|\text{t}0\right\rangle +
d\left|\text{t}1\right\rangle,
\end{equation}
where $a, c, d \in \mathbb{C}$, expanded in the EWF bases is
\begin{equation}\label{31}
\begin{split}
\left|\psi_{C}\right\rangle=&
\frac{a+c+d}{2}\left|\square+\right\rangle +
\frac{a+c-d}{2}\left|\square-\right\rangle +\\&
\frac{a-c-d}{2}\left|\bigcirc+\right\rangle +
\frac{a-c+d}{2}\left|\bigcirc-\right\rangle,
\end{split}
\end{equation} 
\begin{equation}\label{32}
\left|\psi_{AC}\right\rangle=
\frac{a}{\sqrt{2}}\left|\text{h}+\right\rangle +
\frac{a}{\sqrt{2}}\left|\text{h}-\right\rangle +
\frac{c+d}{\sqrt{2}}\left|\text{t}+\right\rangle +
\frac{c-d}{\sqrt{2}}\left|\text{t}-\right\rangle,
\end{equation} 
\begin{equation}\label{33}
\left|\psi_{CB}\right\rangle=
\frac{a+c}{\sqrt{2}}\left|\square0\right\rangle +
\frac{a-c}{\sqrt{2}}\left|\bigcirc0\right\rangle +
\frac{d}{\sqrt{2}}\left|\square1\right\rangle -
\frac{d}{\sqrt{2}}\left|\bigcirc1\right\rangle,
\end{equation} 
where the normalization constraint ($\left\langle\psi|\psi\right\rangle=1$) 
and the surjective isometry constraints: $c - d = 0$ in \eqref{32} to suppress $\left|\text{t}-\right\rangle$, and $a - c = 0$ in \eqref{33} to suppress $\left|\bigcirc0\right\rangle$, imposed to satisfy the conditions $\text{2.}$ and $\text{2'.}$ of the quantum-cakes explanation, lead to the unique solution of $a = c = d$ with probability amplitude $d$ having modulus $\left|d\right| = 1/\sqrt{3}$. Expansions $\left|\psi_{AB\text{th}}\right\rangle$ \eqref{22} and $\left|\psi_{AB\text{ht}}\right\rangle$ \eqref{23} do not represent this solution.

Therefore, the entangled quantum state that Alice delivers to Bob may not be $\left|\psi_{AB}\right\rangle$ \eqref{8} but $\left|\psi_{AB\text{th}}\right\rangle$ \eqref{22} and $\left|\psi_{AB\text{ht}}\right\rangle$ \eqref{23} as well, regardless of the outcome of her measurement of the first qubit \eqref{1}. The contradiction of the EWF cannot be discussed in isolation from Alice's freedom of choice understood as her fallibility.

\section{DISCUSSION}\label{sec:4}
The legitimacy of the EWF procedure has been questioned \cite{11,12}, respectively under a nonunitary account of quantum state reduction, and decoherence. Indeed, the measurement problem is the only source of contradictions in Bell-type experiments. But in the case of the EWF, it is not the incompatibility of the unitary evolution and the measurement involved in the piecewise-defined expression \eqref{2} that is not self-consistent \cite{13,14}, but this nonorthogonal piecewise definition, as such. In this particular one out of twelve (on average) rounds of the experiment where $\left|\bigcirc-\right\rangle$ is measured by Charlie the conditional expression \eqref{2} cannot be guaranteed to hold and possible errors of Alice must be accounted for. The presence of these errors should discourage a casino manager from offering a gambling game based on the principles of the EWF. In a dispute between a gambler and a casino, the judge should rule in favor of the gambler: shifting the responsibility for the erroneous operation of this game to the gambler (consumer) appears to be an unfair commercial practice. Fortunately, such considerations are academic and no judge will ever need to rule in such a case as the EWF is impossible to be implemented as a game in a casino. Unitary transformations of the \emph{big quantum state} \eqref{8} performed within the lab by Alice herself \cite{2} pursuant to the recipes of \eqref{1} and \eqref{2}, or similar, are impossible.

Numerous other publications (e.g., \cite{15,16,17}) attempt to explain the EWF using Bohmian mechanics. But Bohmian interpretation of QT is incompatible \cite{24} with the assumptions of 
\begin{enumerate}
  \item[(i)  ] the choice of which measurement is performed can be made randomly and independently of the system under observation,
  \item[(ii) ] the system has limited memory, and
  \item[(iii)] Landauer's erasure principle holds.
\end{enumerate}  
Assumption (i) is the standard “Freedom of Choice” assumption of Bell-type experiments (freedom of what to measure), while assumptions (ii) and (iii) form the basis of the prevailing explanation of Maxwell demon (Szilard's engine) paradox. The author sees no reason to question these assumptions. “Pilot-wave theories are parallel-universe theories in a state of chronic denial” \cite{25}. Multiverses can be neither experimentally confirmed nor falsified.

Much of the essence of QT already makes itself known in the case of just two nonorthogonal states \cite{26}. But in the case of the EWF, the specific type of quantum states \eqref{1}, \eqref{2}, \eqref{8}, measurements, outcomes and actions involved in the argument are also relevant and should not be omitted \cite{20}. This is important if one compares the EWF with Bell-type experiments, in particular with those belonging to their subset, which excludes the coexistence of observer-independent measurements \cite{19,20,21,22} (termed as Bell-Wigner-type experiments in Ref. \cite{21}). Observer-independent measurements do not exist \cite{19,20} as unitary quantum mechanics invalidates singletime predictions for all observers \cite{27}. Yet, this conclusion cannot be derived through the backdoor. It manifests in collected statistics of measurements of a specific, entangled state, but not in the flawed argument of a superposed-action.

On the contrary to the EWF, a casino manager should not be discouraged from offering a gambling game based on the principles of the Bell-Wigner-type experiments. They are by all means implementable in practice, while errors are relatively small. 6-photon Bell-Wigner-type experiment violated the associated Bell-type inequality by five standard deviations \cite{21}. But here Alice, Bob and Charlie are photon detectors, the activations of which are processed by a classical computer to find 6-photon coincidence events. Thus Alice and Bob not only inform Charlie about obtaining definite measurement results \cite{7} (using heralding signals $\alpha'$ and $\beta'$) but also reveal these results, and yet do not destroy the superposition of the entangled state. As Bell-Wigner-type experiments boil down to nonlocal correlations of observer-dependent measurements, which correlations are known at least since Bells' remarkable theorem, a judge would be presented with an easy task in any dispute between a gambler and a casino: lack of observer-independent measurements \cite{19,20} is a known, experimentally proven \cite{21,22} feature of QT. Different measurement times could be easily introduced in Bell-Wigner-type setups. In the case of photonic implementation, physical delays can be employed on particular light guides between a laser and detectors. On the other hand, in a relativistic frame of reference of a photon, no time passes between the emission of the photon and its absorption, which is otherwise known as time dilation. A quantum state (in particular an entangled one) is time-independent.

Quantum nonlocality reveals that relativistic local causality is only an appearance; measurements and \emph{human free choices} let space-time emerge \cite{28}. In particular, perceived space-time dimensionality, with three spatial dimensions and imaginary time, arises from the exotic $\mathbb{R}^4$ property of such a configuration which is absent in other dimensionalities, and made biological evolution possible \cite{29}. Certainly, Wigner-type experiments are not \emph{miracle} narratives beyond the domain of science \cite{28}.

\section{CONCLUSION}\label{sec:5}
Designers of the EWF argue to have arrived at the contradiction by \emph{letting} the second state \eqref{2}, which Bob receives from Alice, to depend on a random value measured and known by Alice \cite{2}. But since there is no unitary transformation that would bring a pure, single qubit state \eqref{1} and two nonorthogonal versions of a mixed state \eqref{2} into an entangled, pure, two qubit state \eqref{8}, as discussed above, this argumentation is false.

\begin{acknowledgments}
I thank my wife, her mother, and Mirek for their support.
\end{acknowledgments}

\nocite{*}

\bibliographystyle{ieeetr} 
\bibliography{apssamp}

\end{document}